\documentstyle[12pt]{article}
\normalbaselines
\begin{document}
\bibliographystyle{unsrt}

\begin{center}
{{\bf SYMPLECTIC TOMOGRAPHY OF NONCLASSICAL STATES OF TRAPPED ION}}
\end{center}
\begin{center}                       
Olga Man'ko\footnote{On leave from Lebedev Physical Institute, 
Moscow, Russia.\\
Contribution to the Adriatico Research Conference ``Interferometry 2'' 
(ICTP, Trieste, 1--11 March, 1996), originally submitted in English 
13 March, 1996.}\\
{\it International Center for Theoretical Physics, Trieste, Italy}

\end{center}

\begin{abstract} The marginal distribution of squeezed and rotated quadrature
for two types of nonclassical states of trapped ion -- for squeezed and 
correlated states and for squeezed even and odd coherent states (squeezed 
Schr\"odinger cat states) is studied. The obtained marginal distribution 
for the two types of states is shown to satisfy classical dynamical equation 
equivalent to standard quantum evolution equation for density matrix 
(wave function) derived in symplectic tomography scheme.
\end{abstract}
    
\section{Introduction}

\noindent

Recently, in~\cite{Vogel} it was shown that the steady state of trapped
ion irradiated by bichromatic laser field is superposition of two 
coherent states~\cite{Glauber63} which is even/odd coherent state 
introduced in~\cite{Physica74} and interpreted in~\cite{YurkeStoler} 
for large amplitudes of the superposition partners as Schr\"odinger cat 
states~\cite{Schr35}. The theory of ion in the Paul trap was developed 
in~\cite{Glaconf,SchraMa}, where the trapped ion was described by the
model of quantum oscillator with periodically varying frequency.
Gaussian packets, discrete modes and propagator of such oscillator have 
been obtained in~\cite{Husimi}. Linear in position and momentum 
integrals of motion have been found for the oscillator 
in~\cite{Glaconf,SchraMa,MM70}. In~\cite{VogRis}, on the basis 
of analysis~\cite{GlaCah}, the procedure was formulated to obtain the
Wigner function of quantum system in terms of marginal distribution of
rotated quadrature which may be measured by balanced homodyne detector.
The scheme which was called optical tomography has been used 
experimentally~\cite{Raymer}. In~\cite{Mancini1}, the symplectic tomography
procedure has been suggested in which measuring the quantum states
was proposed by means of measuring marginal distribution for squeezed, 
rotated and shifted quadrature. 

In~\cite{Mancini2}, a new equation in quantum mechanics was
introduced describing time evolution of this marginal distribution.
The equation has completely classical form but contains all information 
about quantum system.

The aim of this work is to consider two important types of nonclassical 
states of the trapped ion. First we consider squeezed and correlated 
states~\cite{Kurm} of the ion in the Paul trap~\cite{Glaconf,SchraMa}. 
Then we study also even and odd coherent states of the ion irradiated by 
bichromatic laser field~\cite{Vogel}. The states of trapped ion are 
investigated in frame of symplectic tomography procedure suggested 
in~\cite{Mancini1} and using the new quantum evolution equation 
of~\cite{Mancini2}. It is worthy to note that in~\cite{schleich} 
endoscopy procedure to measure the quantum states of the trapped 
ion is elaborated in detail.  

\section{Gaussian Wigner Function}

\noindent

The generic mixed squeezed state of the trapped ion with the density operator 
$~\hat \rho ~$ is described by the Wigner function $~W\,(p,\,q)~$ of the 
Gaussian form which contains five real parameters~\cite{Olga} (the one-mode 
oscillator case). Two parameters are mean values of momentum 
$~\langle p\rangle~$ and position $~\langle q\rangle~$ and other three 
parameters are matrix elements of the real dispersion matrix $~m\,:$ 
\begin{eqnarray}\label{gwf1}
m_{11}&=&\sigma _{pp}~;\nonumber\\
m_{12}&=&\sigma _{pq}~;\\
m_{22}&=&\sigma _{qq}~.\nonumber
\end{eqnarray}
Below, we will use the invariant parameters
\begin{equation}\label{gwf2}
T=\mbox {Tr}~m=\sigma _{pp}+\sigma _{qq}
\end{equation}
and
\begin{equation}\label{gwf3}
d=\det ~m=\sigma _{pp}\,\sigma _{qq}-\sigma _{pq}^2~.
\end{equation}
For one-mode system, the generic Gaussian Wigner function has the 
form (see, for example,~\cite{Olga,Vol183}\,)
\begin{equation}\label{gwf4}
W\,(p,\,q)=\frac {1}{\sqrt d}~\exp \left \{-\frac {1}{2d}\left [
\sigma _{qq}(p-\langle p\rangle )^{2}+\sigma _{pp}(q-\langle q\rangle )^{2}
-2\sigma _{pq}(p-\langle p\rangle )(q-\langle q\rangle )\right ]\right \}~.
\end{equation}
The parameters $~\langle p\rangle~$ and $~\langle q\rangle~$ are given 
by the formulas
\begin{eqnarray}
\langle p\rangle &=&\mbox {Tr}~\hat \rho \hat p~;\label{gwf5}\\
\langle q\rangle &=&\mbox {Tr}~ \hat \rho \hat q~,\label{gwf6}
\end{eqnarray}
where the operators $~\hat p~$ and $~\hat q~$ are the quadrature 
components of creation $~a^\dagger ~$ and annihilation 
$~a~$~operators
\begin{eqnarray}
\hat p=\frac{a-a^\dagger}{i\sqrt 2}~;\label{gwf7}\\
\hat q=\frac{a+a^\dagger}{\sqrt 2}~.\label{gwf8}
\end{eqnarray}
The matrix elements of the real symmetric dispersion matrix $~m~$ are 
defined as follows
\begin{eqnarray}\label{gwf9}
\sigma _{pp}&=&\mbox {Tr}~\hat \rho \hat p^2-\langle p\rangle ^2~;\nonumber\\
\sigma _{qq}&=&\mbox {Tr}~\hat \rho \hat q^2-\langle q\rangle ^2~;\\
\sigma _{pq}&=&\frac{1}{2} \mbox {Tr}~\hat \rho \,(\hat p\hat q
+\hat q\hat p)-\langle p\rangle \langle q\rangle~.\nonumber
\end{eqnarray}
Due to the physical meaning of the dispersions, the parameters 
$~\sigma _{pp}~$ and $~\sigma _{qq}~$ must be nonnegative numbers,
so the invariant parameter $~T~$ (\ref{gwf2}) is a positive number.
Also the determinant $~d~$ (\ref{gwf3}) of the dispersion matrix must 
be positive. For a pure Gaussian state, the parameter $~d=1/4~.$
   
\section{Squeezed States of Trapped Ion}

\noindent

Since the ion in the Paul trap is described by the model of parametric
oscillator in this section we review its properties. For the parametric 
oscillator with arbitrary time dependence of the frequency and the 
Hamiltonian 
\begin{equation}\label{ss1}
H=-\frac {\partial ^{2}}{2\,\partial x^{2}}
+\frac {\omega ^{2}\,(t)\,x^{2}}{2}~,
\end{equation}
where we put $~\hbar=m=\omega \,(0)=1~$ and used expressions for the position
and momentum operators in the coordinate representation, there exists the 
time-dependent integral of motion found in~\cite{MM70}
\begin{equation}\label{ss2}
A=\frac {i}{\sqrt 2}\,\left [\,\varepsilon \,(t)\,\hat p
-\dot \varepsilon \,(t)\,\hat q\,\right ]~,
\end{equation}
where
\begin{eqnarray}\label{ss3}
\ddot \varepsilon \,(t)+\omega ^{2}\,(t)\,\varepsilon \,(t)&=&0~;\nonumber\\
\varepsilon \,(0)&=&1~;\\
\dot \varepsilon \,(0)&=&i~,\nonumber
\end{eqnarray}
satisfying the commutation relation
\begin{equation}\label{ss4}
[A,\,A^\dagger ]=1~.
\end{equation}
For the trapped ion, the time-dependence of frequency is taken to be 
periodical one~\cite{Glaconf}
\begin{equation}\label{insert1}
\omega ^2 \,(t)=1+\kappa ^2\sin ^2\Omega t~.
\end{equation}
It is easy to show that packet solutions to the Schr\"odinger equation
may be introduced and interpreted as coherent states~\cite{MM70}, since 
they are eigenstates of the operator $~A~$ (\ref{ss2}), of the form
\begin{equation}\label{ss5}
\Psi _{\alpha }\,(x,t)=\Psi _{0}\,(x,t)\exp \left \{
-\frac {|\alpha |^{2}}{2}-
\frac {\alpha ^{2}\varepsilon ^{*}(t)}{2\,\varepsilon \,(t)}
+\frac {{\sqrt 2}\,\alpha \,x}{\varepsilon}\right \}~,
\end{equation}
where
\begin{equation}\label{ss6}
\Psi _{0}\,(x,t)=\pi ^{-1/4}\,[\,\varepsilon \,(t)\,]^{-1/2}
\exp \frac {i\,\dot \varepsilon \,(t)\,x^{2}}{2\,\varepsilon \,(t)}
\end{equation}
is analog of the ground state of the oscillator and $~\alpha ~$ is a
complex number. Variances of the position and momentum of the
parametric oscillator in the state (\ref{ss6}) are
\begin{eqnarray}\label{ss7}
\sigma _{qq}&=&\frac {\mid \varepsilon \,(t)\mid ^{2}}{2}~;\nonumber\\
\\
\sigma _{pp}&=&\frac {\mid \dot \varepsilon \,(t)\mid ^{2}}{2}~,\nonumber
\end{eqnarray}
and the correlation coefficient $~r~$ of the position and momentum has
the value corresponding to minimization of the Schr\"odinger uncertainty
relation~\cite{Schr30}
\begin{eqnarray}\label{ss8}
\sigma _{qq}\,\sigma _{pp}&=&\frac {1}{4}\,\frac {1}{1-r^{2}}~;\nonumber\\
\\
r&=&\frac {\sigma _{pq}}{\sqrt {\sigma _{qq}\,\sigma _{pp}}}~.\nonumber
\end{eqnarray}
If $~\sigma _{qq}<1/2~~(\sigma _{pp}<1/2)~$ we have squeezing in 
quadrature components.

The analogs of orthogonal and complete system of number states which 
are excited states of the ion in the Paul trap are obtained by expansion 
of (\ref{ss5}) into power series  in $~\alpha \,.$ We have
\begin{equation}\label{ss9}
\Psi _{m}\,(x,t)=\left (\frac {\varepsilon ^{*}(t)}{2\,\varepsilon \,(t)}
\right )^{m/2}\frac {1}{\sqrt {m!}}~\Psi _{0}\,(x,t)~H_{m}\left (\frac {x}
{\mid \varepsilon \,(t)\mid }\right ),
\end{equation}
and these squeezed and correlated number states are eigenstates of the
invariant $~A^{\dagger }A~.$

Another normalized solution to the Schr\"odinger equation 
\begin{equation}\label{ss10}
\Psi _{\alpha }^{(+)}\,(x,t)=2\,N^{(+)}~\Psi _{0}\,(x,t)~\exp \left \{
-\frac {\mid \alpha \mid ^{2}}{2}-\frac {\varepsilon ^{*}(t)\,\alpha ^{2}}
{2\,\varepsilon \,(t)}\right \}~
\cosh \frac {{\sqrt 2}\,\alpha \,x}{\varepsilon \,(t)}~,
\end{equation}
where
\begin{equation}\label{ss11}
N^{(+)}=\frac {\exp ~(\mid \alpha \mid ^{2}/2)}
{2\,\sqrt {\cosh \mid \alpha \mid ^{2}}}~,
\end{equation}
is the squeezed even coherent state~\cite{Physica74} (the squeezed
Schr\"odinger cat state). The odd coherent state of the parametric 
oscillator 
\begin{equation}\label{ss12}
\Psi _{\alpha }^{(-)}\,(x,t)=2\,N^{(-)}~\Psi _{0}\,(x,t)~\exp \left \{
-\frac {\mid \alpha \mid ^{2}}{2}
-\frac {\varepsilon ^{*}(t)\,\alpha ^{2}}{2\,\varepsilon \,(t)}\right \}~
\sinh \frac {\sqrt {2}\,\alpha \,x}{\varepsilon \,(t)}~,
\end{equation}
where
\begin{equation}\label{ss13}
N^{(-)}=\frac {\exp ~(\mid \alpha \mid ^{2}/2)}
{2\,\sqrt {\sinh \mid \alpha \mid ^{2}}}~,
\end{equation}
satisfies the Schr\"odinger equation and is the eigenstate of the 
integral of motion $~A^{2}~$ (as well as the even coherent state) 
with the eigenvalue $~\alpha ^{2}\,.$ These states are one-mode examples
of squeezed and correlated Schr\"odinger cat states constructed 
in~\cite{Nikon}.

\section{Tomography of Trapped Ion}

\noindent

In~\cite{Mancini1}, it was shown that for the  generic linear combination
of quadratures which is a measurable observable $~(\,\hbar =1\,)$
\begin{equation}\label{X}
\widehat X=\mu \hat q+\nu\hat p+\delta ~,
\end{equation}
where $~\hat q~$ and $~\hat p~$ are the position and momentum, 
respectively, the marginal distribution
$~w~(X,\,\mu,\,\nu,\,\delta )~$ (normalized with respect to the $~X~$ 
variable), depending upon three extra real parameters
$~\mu;~\nu;~\delta ,\,$ is related to the state of the quantum system 
expressed in terms of its Wigner function $~W(q,\,p)$ as follows
\begin{equation}\label{w}
w\,(X,\,\mu,\,\nu,\,\delta)=\int \exp\,[-ik(X-\mu q-\nu
p-\delta)]~W\,(q,\,p)~\frac{dk~dq~dp}{(2\pi)^2}~.
\end{equation}
As it follows from this formula, the marginal distribution depends
on the difference of the variables 
$$X-\delta =Y~.$$ 
So, we could introduce
\begin{equation}\label{insert2}
P\,(Y,\,\mu ,\,\nu )=w\,(X,\,\mu ,\,\nu ,\,\delta )=w\,(X=Y,\,\mu ,\,\nu ,\,0)~,
\end{equation}
which is marginal distribution depending on three variables.
The physical meaning of the parameters $~\mu;~\nu;~\delta ~$ is that 
they describe ensemble of shifted, rotated and scaled reference frames 
in which the position $~X~$ is measured. Formula (\ref{w}) can be inverted 
and the Wigner function of the state can be expressed in terms of
the marginal distribution~\cite{Mancini1}
\begin{equation}\label{W}
W(q,\,p)=(2\pi)^2s^2\exp \,(isX)~w_F\,(X,\,sq,\,sp,\,s)~,
\end{equation}
where $~w_F\,(X,\,a,\,b,\,s)~$ is the Fourier component of the marginal 
distribution (\ref{w}) taken with respect to the parameters 
$~\mu;~\nu;~\delta ,\,$ namely,
\begin{equation}\label{wF}
w_F\,(X,\,a,\,b,\,s)=\frac{1}{(2\pi)^3}\int w\,(X,\mu,\nu,\delta)~
\exp\,[-i(\mu a+\nu b+\delta s)]~d\mu ~d\nu ~d\delta ~.
\end{equation}
In~\cite{Mancini2}, it was shown that for systems with the Hamiltonians
\begin{equation}\label{HV}
\hat H=\frac{{\hat p}^2}{2}+V(\hat q)~,
\end{equation}
the marginal distribution satisfies the quantum time-evolution equation 
\begin{equation}\label{FPeq}
\dot w-\mu\frac{\partial}{\partial\nu}w-
i\left[V\left(\frac{1}{\partial/\partial\delta}
\frac{\partial}{\partial\mu}+i~\frac{\nu}{2}
\frac{\partial}{\partial\delta}\right)-
V\left(\frac{1}{\partial/\partial\delta}
\frac{\partial}{\partial\mu}-i~\frac{\nu}{2}
\frac{\partial}{\partial\delta}\right)\right]w=0~.
\end{equation}
The dot implies partial time derivative.
This equation can be rewritten for the function $~P~$ (\ref{insert2})
in the form
\begin{equation}\label{insert3}
\dot P-\mu\frac{\partial}{\partial\nu}P-
i\left[V\left(\frac{-1}{\partial/\partial Y}
\frac{\partial}{\partial\mu}-i~\frac{\nu}{2}
\frac{\partial}{\partial Y}\right)-
V\left(\frac{-1}{\partial/\partial Y}
\frac{\partial}{\partial\mu}+i~\frac{\nu}{2}
\frac{\partial}{\partial Y}\right)\right]P=0~.
\end{equation}
The measurable position $~Y~$ is cyclic variable for the evolution equation.
For the trapped ion, Eq.~(\ref{FPeq}) takes the form
\begin{equation}\label{TIE}
\dot w-\mu \,\frac{\partial}{\partial\nu}\,w+\omega ^2\,(t)\,\nu \,
\frac{\partial}{\partial\mu}\,w=0~,
\end{equation}
or in terms of the function $~P,$
\begin{equation}\label{insert4}
\dot P-\mu \,\frac{\partial}{\partial\nu}\,P+\omega ^2\,(t)\,\nu \,
\frac{\partial}{\partial\mu}\,P=0~.
\end{equation}
For squeezed and correlated states of the trapped ion, the Wigner function
has the form (\ref{gwf4}) in which quadrature variances and covariance
for the state ({\ref{ss6}) are given by ({\ref{ss7}) and ({\ref{ss8}) 
and quadrature means are equal to zero. For the state ({\ref{ss5}),
the dispersion matrix is the same but two quadrature means have
nonzero values
\begin{eqnarray}
\langle p\rangle &=&\frac {1}{\sqrt 2}\left(\alpha \dot \varepsilon ^*
+\alpha ^*\dot \varepsilon \right)~;\label{insert5a}\\
\langle q\rangle&=&\frac {1}{\sqrt 2}\left(\alpha \varepsilon ^*
+\alpha ^*\varepsilon \right)~.\label{insert5b}   
\end{eqnarray}
Calculating integral (\ref{w}) one can show that for generic
Gaussian packets of the trapped ion [also, for particular cases
(\ref{ss5}) and (\ref{ss6})\,] the marginal distribution is
\begin{equation}\label{freesolution}
w\,(X,\,\mu,\,\nu,\,\delta ,\,t)=
\frac{1}{\sqrt{2\,\pi \,\sigma_X(t)}}\,\exp\left\{-\frac{(X-\bar {X})^2}
{2\,\sigma _X(t)}\right\}~,
\end{equation}
where the dispersion of the symplectic observable $~X~$ and the mean
value of the observable depend on time and parameters as follows
\begin{eqnarray}
\sigma _X(t)&=&\mu ^2\sigma _{qq}+\nu ^2\sigma _{pp}
+2\,\mu \,\nu \,\sigma _{pg}~;\label{insert6}\\
\bar {X}&=&\mu \langle q\rangle +\nu \langle p\rangle 
+\delta ~.\label{insert7}
\end{eqnarray}
It is the same form of Gaussian distribution discussed for free motion
and harmonic oscillator in~\cite{Mancini2} but with different 
quadrature dispersions and means which are given by (\ref{ss7}),
(\ref{ss8}) and (\ref{insert5a}), (\ref{insert5b}), correspondingly.

One can check that the normalized marginal distribution 
(\ref{freesolution}) with parameters (\ref{insert6}) and (\ref{insert7})
and quadrature dispersions and means (\ref{ss7}), (\ref{ss8}) and
(\ref{insert5a}), (\ref{insert5b}) satisfies the evolution equation
(\ref{TIE}). This follows from the observation that the marginal
distribution depends on the symplectic parameters and time only through
the dependence on the mean value $~\bar X~$ and dispersion 
$~\sigma _X(t)~.$ Then calculating the first derivatives of the 
marginal distribution in all variables and using the Ehrenfest theorem for
quadrature means and also the evolution equations for variances and 
covariance of the parametric oscillator~\cite{Vol183}, one can show that
the evolution equation is satisfied by the function (\ref{freesolution}).

The evolution of the Wigner function of trapped ion (as the system  
with quadratic Hamiltonian) for any state is given by 
the following prescription (see, for example,~\cite{el}\,). 
Given the Wigner function $~W\,(p,\,q,\,t=0)$
at the initial moment of time $~t=0.$ Then the Wigner function at time 
$~t$ is obtained by the replacement
$$W\,(p,\,q,\,t)=~W\,[\,p\,(t),\,q\,(t),\,t=0\,]~,$$
where time-dependent arguments are the linear integrals of motion
of the quadratic system. The linear integrals of motion describe 
initial values of classical trajectories in the phase space of the 
system
\begin{eqnarray}\label{xxx}
p\,(t)&=&\frac {\varepsilon +\varepsilon ^*}{2}\,p-
\frac {\dot \varepsilon +\dot \varepsilon ^*}{2}\,q~;\nonumber\\
\\
q\,(t)&=&\frac {\varepsilon -\varepsilon ^*}{2i}\,p+
\frac {\dot \varepsilon -\dot \varepsilon ^*}{2i}\,q~.\nonumber
\end{eqnarray}

This ansatz follows from the statement that the density 
operator of the Hamiltonian system is the integral of motion and 
its matrix elements in any basis must depend on appropriate integrals 
of motion. Consequently, for arbitrary mixed state of the ion in the
Paul trap with the initial Gaussian Wigner function~(\ref{gwf4}), time
evolution of the Wigner function is given by the same formula~(\ref{gwf4}) 
in which the variables $~p~$ and $~q~$ are replaced by the 
functions~(\ref{xxx}), correspondingly. 
                
The marginal distribution is related to the
Wigner quasi-distribution function by (\ref{w}). So, one can extend 
the above anzats to the marginal distribution. We formulate this 
anzats for the Paul trap as follows. Given an initial distribution
function
\begin{equation}\label{x}
w\,(X,\,\mu ,\,\nu ,\,\delta ,\,t=0)=w_0\,(X-\delta ,\,\mu ,\,\nu )~.
\end{equation}
Then at time $~t\neq 0,$ the marginal distribution which is the
solution to Eq.~(\ref{TIE}) has the form
\begin{equation}\label{xx}
w\,(X,\,\mu ,\,\nu ,\,\delta ,\,t)=w_0\,(\,X-\delta ,\,\mu (t),\,\nu (t)\,)~,
\end{equation}
in which                                  
\begin{eqnarray}\label{xxxx}     
\nu \,(t)&=&\frac {\dot \varepsilon -\dot \varepsilon ^*}{2i}\,\nu
+\frac {\varepsilon -\varepsilon ^*}{2i}\,\mu ~;\nonumber\\
\\
\mu \,(t)&=&\frac {\dot \varepsilon +\dot \varepsilon ^*}{2}\,\nu
+\frac {\varepsilon +\varepsilon ^*}{2}\,\mu ~.\nonumber
\end{eqnarray}
                                          
\section{Even and Odd Coherent States}

\noindent

Now we will discuss marginal probability for nonclassical states of the 
parametric oscillator, such as even and odd coherent states~\cite{Physica74}.
To describe such nonclassical states, as even and odd coherent states 
not only for an ion in the Paul trap but also for other types of trapped 
ions we consider  construction of multimode even and odd coherent 
states (see, for example,~\cite{el,ans}\,). Symplectic tomography
procedure~\cite{Mancini1} was extended for multimode systems 
in~\cite{Ariano}.

We define multimode even and odd coherent states as
\begin{equation}
\mid \mbox {\bf A}_{\pm}\rangle =N^{(\pm )}\left (\,\mid 
\mbox {\bf A}\rangle \pm \mid -\mbox {\bf A}\rangle \,\right)~,
\label{e23}
\end{equation}
where the multimode coherent state $~\mid \mbox {\bf A}\rangle ~$ 
is
\begin{equation}
\mid \mbox {\bf A}\rangle =\mid \alpha_{1},~\alpha_{2},~\ldots ,
~\alpha_{n}\rangle =D(\mbox {\bf A}) \mid \mbox {\bf 0}\rangle ~,
\label{e24}
\end{equation}
and the multimode coherent state is created from the multimode vacuum 
state $~\mid \mbox {\bf 0}\rangle ~$ by the multimode displacement 
operator $~D\,(\mbox {\bf A}).$
The definition of multimode even and odd coherent states is the 
obvious generalization of the one-mode even and odd coherent states. 
Normalization constants for multimode even and odd coherent states 
are
\begin{eqnarray}\label{e25} 
N^{(+)}&=&\frac {\exp \,(\mid \mbox {\bf A}\mid^{2}/2)}
{2\,\sqrt{\cosh\mid \mbox {\bf A}\mid^{2}}}~;\nonumber\\
\\
N^{(-)}&=&\frac {\exp \,(\mid \mbox {\bf A}\mid^{2}/2)}
{2\,\sqrt {\sinh \mid \mbox {\bf A}\mid^{2}}}~,\nonumber
\end{eqnarray}
where $~\mbox {\bf A}=(\alpha_{1},~\alpha_{2},~\ldots,~\alpha_{n})~$ 
is a complex vector.

The Wigner function for multimode coherent states is (see~\cite{Vol183}\,) 
\begin{equation}
W_{\mbox {\bf A,B}}
=2^{N}\,\exp \left (-2\,\mbox{\bf ZZ}^{*}+2\,
\mbox {\bf AZ}^{*}+2\,\mbox {\bf B}^{*}\mbox {\bf Z}
-\mbox {\bf AB}^{*}-\frac {\mid \mbox {\bf A}\mid^{2}}{2}
-\frac {\mid \mbox {\bf B}\mid^{2}}{2}\right )~,
\label{e38}
\end{equation}
where
\begin{equation}
\mbox {\bf Z}=\frac {\mbox {\bf q}+i\mbox {\bf p}}{\sqrt 2}~.
\label{e39}
\end{equation}
For even and odd coherent states, the Wigner  function is (see~\cite{ans}\,)
\begin{eqnarray}
W_{\mbox {\bf A}_{\pm}}\,(\mbox {\bf q,\,p})
&=&\mid N^{(\pm )}\mid^{2}\left [\right. W_{\mbox {\bf (A,B=A)}}\,
(\mbox {\bf q,\,p})\pm W_{\mbox {\bf (A,B=-A)}}\,(\mbox {\bf q,\,p})
\nonumber\\
\nonumber\\
&\pm &W_{\mbox {\bf (-A,B=A)}}\,(\mbox {\bf q,\,p})+
W_{\mbox {\bf (-A,B=-A)}}\,(\mbox {\bf q,\,p})\left. \right ]~,
\label{e40}
\end{eqnarray}
where explicit forms of $~N^{(\pm )}~$ are given by Eq.~(\ref{e25}).
For multimode case, we use following notation
\begin{eqnarray}
\mbox {\bf AZ}^{*}&=&\alpha_{1}Z_{1}^{*}+\alpha_{2}Z_{2}^{*}
+\cdots \alpha_{n}Z_{n}^{*}~;\nonumber\\
\\
\mbox {\bf ZZ}^{*}&=&Z_{1}Z_{1}^{*}+Z_{2}Z_{2}^{*}
+\cdots +Z_{n}Z_{n}^{*}~.\nonumber
\label{e41}
\end{eqnarray}
If at time $~t=~0,$ one has the initial Wigner function of the system
in the form
\begin{equation}\label{gq5}
W\,(\mbox {\bf p},\mbox {\bf q},t=0)=W_{0}\,(\mbox {\bf Q})~,
\end{equation}
the Wigner function of the system at time $~t~$ is
\begin{equation}\label{gq6}
W\,(\mbox {\bf p},\,\mbox {\bf q},\,t)=W_{0}\,[\,\Lambda \,(t)\,\mbox {\bf Q}
+\Delta \,(t)\,].
\end{equation}
where the matrix $~\Lambda \,(t)~$ and the vector 
$~\Delta \,(t)~$ 
determine the linear integrals of motion for quadratic systems 
to which the parametric oscillator belongs~\cite{el}.

Thus, we get the expressions for the Wigner functions of even and odd coherent 
states of the trapped ion, if in formulas~(\ref{e40}) we put $~n=1~$ and 
replace $~\mbox {\bf p}~$ and $~\mbox {\bf q}~$ by the linear integrals 
of motion of the parametric oscillator
\begin{eqnarray}\label{gq16}
 p_0&=&\frac {A-A^\dagger }{i\sqrt 2}~;\nonumber\\
\\
 q_0&=&\frac {A+A^\dagger }{\sqrt 2}~,\nonumber
\end{eqnarray}
where the invariant $~A~$ is given by~(\ref{ss2}).
The explicit form of the invariants $~p_0;~q_0~$ which should be
considered as $~c$--numbers is given by (\ref{xxx}).
Since the marginal distribution is related to density matrix
which is the integral of motion, the time dependence of the marginal
distribution for the parametric oscillator may be given by analogous 
prescription which was used for the Wigner function. If one knows the 
marginal distribution at the initial moment of time as the function of 
the parameters $~\mu ~$ and $~\nu ,$ at time~$~t~$ these parameters 
in the expression for the initial marginal distribution should be 
replaced by linear combinations of $~\mu ~$ and $~\nu $~(\ref{xxxx}).
 
The marginal distribution for initial even and odd coherent states 
was given in~\cite{Mancini2}                       
\begin{equation}\label{xxxxx}   
w^{\pm}\,(\,X,\,\mu ,\,\nu ,\,\delta \,)
=N^{(\pm )2}\,\frac {1}{\sqrt {\pi \left (\mu ^2
+\nu ^2\right )}}\left \{w_1+w_2\pm w_3\pm w_4\right \}~,
\end{equation}
where
\begin{eqnarray*}
w_1&=&\exp \left [-\frac {\left (Y-s-s^*\right )^2}{\mu ^2
+\nu ^2}\right ]~;\nonumber\\
w_2&=&\exp \left [-\frac {\left (Y+s+s^*\right )^2}{\mu ^2
+\nu ^2}\right ]~;\nonumber\\
w_3\,=\,w_4^*&=&\exp \left [-2\mid \alpha \mid ^2-\frac {\left (Y-s
+s^*\right )^2}{\mu ^2+\nu ^2}\right ]~;\nonumber\\
Y&=&X-\delta ~;\nonumber\\
s&=&\sqrt 2\,\alpha \,(\mu -i\nu )~.\nonumber
\end{eqnarray*}
Thus, replacing the parameters $~\nu ~$ and $~\mu ~$ in (\ref{xxxxx})
by time-dependent functions (\ref{xxxx}) we get the marginal distribution 
of the trapped ion at any time moment $~t.$
                                                       
The optical tomography procedure suggested in~\cite{VogRis,Raymer}
is the partial case of symplectic tomography, since for the optical
tomography one has to take the symplectic parameters in the form
\begin{eqnarray}\label{z}
\mu &=&\cos \phi ~;\nonumber\\
\nu &=&\sin \phi ~;\\
\delta &=&0~.\nonumber
\end{eqnarray}
Thus, the marginal distribution for rotated quadrature in case of
initial even and odd coherent states takes the form
\begin{equation}\label{zz}
w^{\pm}\,(\,X,\,\varphi ,\,t=0\,)
=N^{(\pm )2}\,\frac {1}{\sqrt \pi }\,\left \{w_1\,(\,X,\,\varphi \,)
+w_2\,(\,X,\,\varphi \,)\pm w_3\,(\,X,\,\varphi \,)
\pm w_4\,(\,X,\,\varphi \,)\right \}~,
\end{equation}
where
\begin{eqnarray*}
w_1\,(\,X,\,\varphi \,)&=&\exp \left [-\left (X-2\,\sqrt 2\mid \alpha \mid 
\cos \,(\varphi _\alpha -\varphi )\right )^2\right ]~;\nonumber\\
w_2\,(\,X,\,\varphi \,)&=&\exp \left [-\left (X+2\,\sqrt 2\mid \alpha \mid 
\cos \,(\varphi _\alpha -\varphi )\right )^2\right ]~;\nonumber\\    
w_3\,(\,X,\,\varphi \,)\,=\,w_4^*\,(\,X,\,\varphi \,)&
=&\exp \left [-2\mid \alpha \mid ^2-\left (X-i\,2\,\sqrt 2\,
\sin \,(\varphi _\alpha -\varphi )\right )^2\right ]~;\nonumber\\    
\alpha &=&\mid \alpha \mid \,\exp\,i\varphi _\alpha ~.\nonumber
\end{eqnarray*}
This marginal distribution, used in the optical tomography scheme, at 
time $~t\neq 0~$ is given by formula (\ref{xxxxx}) in which the 
variables $~\nu ~$ and $~\mu ~$ are replaced by the variables
\begin{eqnarray}\label{zzz}
\nu \,(t,\,\varphi )&=&\frac {\varepsilon -\varepsilon ^*}{2i}
\,\cos \,\varphi +\frac {\dot \varepsilon 
-\dot \varepsilon ^*}{2i}\,\sin \,\varphi ~;\nonumber\\
\\
\mu \,(t,\,\varphi )&=&\frac {\varepsilon +\varepsilon ^*}{2}\,\cos \,\varphi
+\frac {\dot \varepsilon +\dot \varepsilon ^*}{2}\,\sin \,\varphi ~.\nonumber
\end{eqnarray}
Using (\ref{zz}) and (\ref{zzz}) and measuring the marginal distribution of
homodyne output the Wigner function of the trapped ion may be reconstructed 
by means of the Radon transform~\cite{VogRis}.
                                                                                        
Thus, the rule formulated gives the time dependence of both the marginal 
distribution of rotated quadrature and the marginal distribution of 
squeezed and rotated quadrature for even and odd coherent states of 
trapped ion. Replacing $~\mu ~$ and $~\nu ~$ in (\ref{xxxxx})
by $~\mu \,(t)~$ and $~\nu \,(t)~$ from (\ref{xxxx}),
we get solution to the dynamical equation (\ref{FPeq}) for the ion 
in the Paul trap.

\section{Conclusion}

\noindent

In the present work, we calculated the marginal distribution of
symplectic observable (which is generic linear quadrature) for 
nonclassical states of trapped ion modelled by the parametric 
quantum oscillator. Measurements of the marginal distribution
give the possibility to measure the quantum states. In the case 
of particular choice of the parameters 
$~\mu =\cos \varphi ;~\nu =\sin \varphi ,$                  
the measurement is reduced to finding marginal distribution for 
homodyne output and reconstructing the Wigner function through the 
Radon transform of optical tomography scheme~\cite{VogRis,Raymer}. 
The found distribution of generic linear quadrature satisfies the new
classical-like equation of quantum dynamics introduced in the
symplectic tomography scheme.

\section*{\bf Acknowledgments}

\noindent

This work has been partially supported by the Russian Basic Research 
Foundation.

The author would like to acknowledge the International Center for 
Theoretical Physics in Trieste for kind hospitality.

The author thanks Prof. W. Schleich for discussions and for communication
of the results of~\cite{schleich} before publication.
The  author thanks Prof. V. I. Man'ko and Prof. P. Tombesi for fruitfull
discussions.

\end{document}